\documentstyle[11pt,epsfig,twoside]{article}
\newcommand{\mr}[1]{{\mathrm {#1}}}

\newcommand{\ase}{\alpha_{s}}

\newcommand{\bbar}[1]{$\overline{\mr{#1}}$}
\newcommand{\bbare}[1]{\overline{\mr{#1}}}

\newcommand{\reff}[1]{(\ref{#1})}

\newcommand{\be}{\begin{equation}}
\newcommand{\ee}{\end{equation}}
\newcommand{\bd}{\begin{description}}
\newcommand{\ed}{\end{description}}
\newcommand{\bmat}{\begin{displaymath}}
\newcommand{\emat}{\end{displaymath}}
\newcommand{\bit}{\begin{itemize}}
\newcommand{\eit}{\end{itemize}}
\newcommand{\ben}{\begin{enumerate}}
\newcommand{\een}{\end{enumerate}}

\begin{document}
\begin{flushright}
PRA-HEP/95-04  \\ May 1995
\end{flushright}
\vspace*{0.3cm}

\begin{center}
{\Large \bf On the BLM scale--fixing procedure, its generalizations
and the ``genuine'' higher order corrections}
\\

\vspace*{0.5cm}
{\large Ji\v{r}\'{\i} Ch\'{y}la}\\
{Institute of Physics, Academy of Sciences of the Czech
Republic \\ Na Slovance 2, 18040 Prague 8, Czech Republic}
\end{center}

\vspace*{0.5cm}
\begin{center}
{\large Abstract}
\end{center}

\vspace*{0.1cm}
\noindent
The question of the uniqueness of the Brodsky--Lepage--Mackenzie
procedure for fixing the renormalization scale in perturbative QCD
is discussed. It is shown that the resulting finite order approximants
are as ambiguous as the original truncated perturbative expansions. This
inherent ambiguity of the BLM procedure undermines the recent attempts
to define ``genuine'' higher order perturbative corrections.

\vspace*{0.5cm}
\section{Introduction}
Over the last 15 years the problem of the renormalization prescription
dependence of finite order approximants to perturbation expansions in
QCD
\footnote{In this article only the renormalization prescription
dependence of physical quantities in QCD with $n_f$ massless quark
flavours will be considered.}
has been a subject of lively and sometimes even heated debate.
{}From time to time a ``resolution'' of this problem is announced, but
invariably it turns out that these ``solutions'' contain the original
ambiguity in some disguise.  One of such methods has been
suggested some years ago by Brodsky, Lepage and Mackenzie
(BLM) \cite{BLM}.
Soon after the appearance of the preprint version of \cite{BLM}
Stevenson and Celmaster \cite{CS} pointed out its inherent
ambiguity, but their criticism has largely been ignored and the BLM
procedure has since been used in many phenomenological analyses.
In the journal version of their work the authors of \cite{BLM}
have acknowledged the presence of the ambiguity pointed out in
\cite{CS}, but claimed that it can ``be eliminated to a large extent
by adopting some physical process as a theoretical standard for
defining $\alpha_s(Q)$.'' As the BLM procedure is based on the
generalization (called ``naive nonabelianization'' \cite{BB}) of the
QED procedure of incorporating the effects of
quark loops in the renormalized coupling constant (couplant for
short), it has been conjectured to be closely related to
the large order
behaviour of the expansion coefficients in perturbation theory.
This observation has recently been exploited in several
attempts \cite{BB,BBB,Neubert} to help answer the question of the
importance of higher order corrections in QCD perturbative
expansions. The idea suggested in these papers
is to distinguish the ``genuine'' higher order
corrections from those governed by the choice of the renormalization
scale and, supposedly, related to the large order behaviour of
perturbative expansions.  As this question is
of principal as well as practical importance, it is certainly
useful to reappraise the old criticism voiced in \cite{CS} in order
to see how much of it is relevant for these recent efforts. The main
purpose of this paper is to carry out such an analysis.

The paper is organized as follows. After introducing in Section 2
the notation and recalling some basic facts concerning the description
of ambiguities of finite order perturbative approximants, I shall
discuss in Section 3 the BLM procedure in QCD. This is followed
in Section 4 by a short detour to QED, in order to see why there the BLM
procedure does, indeed, lead to a unique result.
In Section 5 the abovementioned suggestion \cite{BLM} to avoid the
``residual'' RS dependence of the BLM results by selecting some
physical process as a standard for the definition of $\alpha_s$
is shown to be no real cure to the problem.
The implications of the inherent ambiguity of the BLM procedure for
the basic strategy of the papers \cite{BB,BBB,Neubert} are discussed
in Section 6, followed by a short summary and conclusions in
Section 7.

\section{Ambiguities of finite order approximants in perturbative QCD}
Before coming to the essence of the BLM procedure, let me introduce
notation and recall a few basic facts concerning the renormalization
prescription ambiguity of finite order perturbative approximants.
I shall make a clear difference between the  concepts of
renormalization prescription, renormalization scheme and
renormalization convention, which all appear in the literature
and which are not always
used in the same sense. I shall emphasize the mathematically complete
quantitative description of these concepts as this will be crucial for
the later discussion. Let us start by
considering perturbation expansion for the generic physical
quantity $r(Q)$, depending, for simplicity, on a single external
momentum $Q$,
\be
r(Q)= a(\mu,\mr{RS})\left[1+r_{1}(Q/\mu,\mr{RS})a(\mu,\mr{RS})+
 r_{2}(Q/\mu,\mr{RS})a^2(\mu,\mr{RS})+\cdots\right].
\label{r(Q)}
\ee
The generalization of the following considerations for the
case when the leading term behaves as $a^P$ is trivial.
In the above expression
$a(\mu,\mr{RS})\equiv \ase/\pi$ is the {\em renormalized couplant}
(the adjective ``renormalized'' will be dropped in the following),
evaluated at the renormalization scale $\mu$ in a given
{\em renormalization scheme} (RS), to be fully specified below. While
the whole sum in \reff{r(Q)} is independent of both the scale $\mu$ and
the RS, any finite order approximant
\be
r^{(N)}(Q/\mu,\mr{RS})\equiv \sum_{k=0}^{N-1}r_{k}(Q/\mu,\mr{RS})a^{k+1}
(\mu,\mr{RS}),
\label{kon}
\ee
{\em does} depend on $\mu$ as well as the RS. The dependence of the
couplant on the scale $\mu$ is governed by the equation
\be
\frac{\mr{d}a(\mu,\mr{RS})}{\mr{d}\ln\mu}\equiv \beta(a)=
-ba^2(\mu,\mr{RS})
\left(1+ca(\mu,\mr{RS})+c_{2}a^2(\mu,\mr{RS})+\cdots\right).
\label{RG}
\ee
In massless QCD the first two coefficients on the r.h.s. of
\reff{RG} are unique functions of $n_f$, the number of massless
quarks  ($b=(33-2n_{f})/6$, $c=(153-19n_{f})/(33-4n_f)$), while
all the higher order ones are {\em completely arbitrary}
\footnote{These coefficients define the so called
{\em renormalization convention} (RC), introduced in \cite{PMS}.}.
Once they are chosen and some initial condition on $a$ is specified,
\reff{RG} can be solved. One way of specifying this boundary
condition is via the scale parameter $\Lambda_{\mr{RS}}$,
introduced in the following implicit equation for the solution of
\reff{RG} \cite{PMS}
\footnote{
The parameter $\Lambda$ introduced in \reff{okraj}
differs from $\Lambda$ used in most phenomenological
analyses by multiplicative $n_f$-dependent factor
$\left(2c/b\right)^{c/b}$, which for realistic values of $n_f$
is close to unity.}
\be
b\ln\frac{\mu}{\Lambda_{\mr{RS}}}=\frac{1}{a}+c\ln\frac{ca}{1+ca}+
\int^{a}_{0}dx\left[-\frac{1}{x^2 B^{(n)}(x)}+\frac{1}
{x^2 (1+cx)}\right]
\label{okraj}
\ee
where
$B^{(n)}(x)\equiv (1+cx+c_2 x^2 + \cdots +c_{n-1}x^{n-1})$.
The dependence of the couplant on the
parameters $c_i$ for $i\ge 2$ is determined by equations similar to
\reff{RG}
\cite{PMS}
\be
\frac{\mr{d}a(\mu,c_{i})}{\mr{d}c_{i}} \equiv \beta_{i}=-\beta(a)
\int_{0}^{a}\frac{bx^{i+2}}{(\beta(x))^2}dx,
\label{betai}
\ee
which are uniquely determined by the basic $\beta$--function in
\reff{RG} and thus introduce no additional ambiguity. The
renormalization scheme is defined by choosing both the coefficients
$c_i$ and the parameter $\Lambda_{\mr{RS}}$, i.e. by the full
specification of the solution of \reff{RG},
RS$\equiv\{c_i,\Lambda_{\mr{RS}}\}$. Note that at the NLO only
$\Lambda_{\mr{RS}}$ labels the RS. Finally the {\em renormalization
prescription} is defined by the specification of both the
renormalization scale $\mu$ and renormalization scheme RS.
Only the specification of the renormalization
prescription thus leads to unique results of finite order perturbative
approximants \reff{kon}.

In connection with $\Lambda_{\mr{RS}}$ the following fact is worth
mentioning. The BLM procedure, to be discussed in the next section,
is based on the isolation of the $n_f$--dependence of the expansion
coefficients $r_k$. This brings up the subtle question of the
$n_f$--dependence of the chosen RS, which up to the NLO means the
$n_f$--dependence of $\Lambda_{\mr{RS}}$. As worlds with different
number of massless quarks are not related by any theoretical arguments,
such as the renormalization group (RG) considerations, there is,
however, no meaningful way of introducing this dependence. What can,
however, be
done in a well--defined manner is to discuss the $n_f$--dependence of
the {\em ratio} $\Lambda_{\mr{RS}'}/\Lambda_{\mr{RS}}$ of the
$\Lambda$--parameters, corresponding to two different RS, as this ratio
is determined solely by the RG considerations.

After the arbitrary coefficients $c_i$ are specified there
are thus two parameters to vary: $\mu$ and $\Lambda_{\mr{RS}}$. As,
however, $\mu$ enters the solution of \reff{RG} always in the ratio
with $\Lambda_{\mr{RS}}$, we can
\bit
\item select one $\Lambda_{\mr{RS}}$
and thus one of the RS=\{$c_i,\Lambda_{\mr{RS}}$\}
and vary $\mu$ only,
\item fix $\mu$ by identifying it with some external momentum, such as
$Q$, and vary $\Lambda_{\mr{RS}}$ instead.
\eit
Both of these options are completely equivalent and it is merely a
matter of taste which one to use. To vary both the scale $\mu$ and
the $\Lambda_{\mr{RS}}$ is legal, but
obviously redundant. Fixing the scale $\mu$ without also specifying
the RS is, on the other hand, insufficient to specify the
renormalization prescription and thus does not lead to a unique
result for the finite order approximants \reff{kon}.

I emphasize this point as the scale $\mu$ is often fixed by identifying
it with some natural physical scale of the process, such as $Q$.
Although such a natural scale can usually be found, its
mere existence {\em does not} fix the RS and thus does not specify the
couplant. In most of phenomenological analyses
the $\overline{\mr{MS}}$ RS is tacitly adopted, but there is no
theoretical argument for this choice, except that in this RS
and for the conventional choices of $\mu=Q$, the
coefficients $r_k$ are often, but not always, small.  If, however, the
magnitude of the expansion coefficients $r_k$
should be the criterion for the ``best'' choice of the scale
$\mu$ and the RS, we would be naturally drawn to the effective charges
approach \cite{ECH}, where all higher order coefficients $r_k$ vanish by
definition.

While the explicit dependence of the couplant on the renormalization
scale $\mu$ and the parameters $\Lambda_{\mr{RS}},c_i,i\ge 2$
is given by \reff{RG} and \reff{betai}, the
dependence of the coefficients $r_k$ on them is determined by the
relations
\be
\frac{\mr{d}r^{(N)}(Q/\mu,\mr{RS})} {\mr{d}\ln \mu}={\cal
O}(a^{N+1}),\;\;\;\; \frac{\mr{d}r^{(N)}(Q/\mu,\mr{RS})}
{\mr{d}\ln \Lambda_{\mr{RS}}}={\cal O}(a^{N+1}), \;\;\;\;
\frac{\mr{d}r^{(N)}(Q/\mu,\mr{RS})} {\mr{d}c_{i}}={\cal O}(a^{N+1}),
\label{konpod}
\ee
which express the internal consistency of perturbation theory.
Iterating these equations we easily find
\begin{eqnarray}
r_{1}(Q/\mu,\mr{RS}) & = & b\ln\frac{\mu}
{\Lambda_{\mr{RS}}}-\rho(Q/\Lambda_{\mr{RS}}),  \nonumber \\
r_{2}(Q/\mu,\mr{RS}) & = & \rho_{2}-c_{2}+r_{1}^{2}+cr_{1},
\label{r1r2} \\
  & \vdots & \nonumber \\
r_{n}(Q/\mu,\mr{RS}) & = & \rho_{n}-c_{n} + f(r_i,c_i,\rho_i;i\le n-1).
\nonumber
\end{eqnarray}
where the quantities $\rho, \rho_i$ are RG invariants.
The dependence of the perturbative approximants \reff{kon} on
$Q$ comes entirely through the invariant $\rho(Q/\Lambda)$, which
can be written as
\footnote{Despite the appearance of $\Lambda_{\mr{RS}}$ in this
expression, $\rho$ is actually RS--independent as the RS--dependences of
the two terms on the r.h.s. of \reff{rho} mutually cancel.}
\be
\rho=b\ln(Q/\Lambda_{\mr{RS}})-r_1(1,\mr{RS}),
\label{rho}
\ee
while all the higher order invariants $\rho_i$ are just pure numbers.
A nontrivial part of any perturbative calculation beyond the LO
boils down to the evaluation of the invariants $\rho_i$,
the rest being essentially
a straightforward exploitation of the RG considerations based on
(3,5,7).

In the following section the BLM procedure will be discussed in detail
at the NLO, as this is where the dependence on the scale $\mu$
comes in  and where also its basic ambiguity becomes
evident. At this order only the first two, unique, coefficients
$b$ and $c$ in \reff{RG} are taken into account and the renormalization
prescription dependence of \reff{r(Q)} is therefore essentially
a one--parameter ambiguity. Varying either
$\mu$ for fixed $ \Lambda_{\mr{RS}}$, or vice versa, we find
\begin{eqnarray}
a(\mu',\mr{RS}) & = &
a(\mu,\mr{RS})\left(1-b\ln
\left(\frac{\mu'}{\mu}\right)a(\mu,\mr{RS}) +\cdots\right),
\label{change1} \\
a(\mu,\mr{RS'}) & = & a(\mu,\mr{RS})\left(1-b\ln
\left(\frac{\Lambda_{\mr{RS}}}{\Lambda_{\mr{RS'}}}\right)
a(\mu,\mr{RS}).
+\cdots\right)
\label{change2}
\end{eqnarray}
In both ways of labelling the ambiguity, we have
\be
a' = a\left(1-\kappa a+\cdots\right),\;\;\;
r'_{\!1} = r_1 + \kappa
\label{change}
\ee
At the NLO and for some ``initial'' $a\equiv a(\mu,\mr{RS})$
$\kappa$ can thus be used as yet another way of labelling this
ambiguity.  After selecting
\begin{description}
\item {a)} the scale $\mu$ in a fixed RS, or
\item {b)} the parameter $\Lambda_{\mr{RS}}$ for fixed $\mu$, or
\item {c)} the parameter $\kappa$ for a fixed initial $a$ in
\reff{change}
\end{description}
we should, however, check that the resulting couplant and expansion
coefficient $r_1$ {\em do not} depend on the RS (for a)),
$\mu$ (for b)) or initial $a$ (for c)). At the NLO
and in all three ways of labelling the inherent one--parameter
ambiguity, both the {\em Principle of Minimal Sensitivity} (PMS)
\cite{PMS} and {\em Effective Charges} (ECH) \cite{ECH} approaches do,
indeed, lead to a unique result for $r^{(2)}(Q)$ \footnote{Starting at
the NNLO, there are certain complications with the existence and/or
uniqueness of the PMS and ECH ``optimized'' results, but this is
irrelevant for the present discussion.}. On the other hand, as will
be shown in detail in the next section, this basic requirement is not
satisfied by the BLM procedure.

Although the scale $\mu$ appears
naturally and unavoidably in the process of renormalization,
perturbation expansions for {\em physical} quantities can actually do
without it. Combining eqs. \reff{okraj} and \reff{r1r2} allows us to
express all the expansion coefficients $r_k$ as unique functions of the
$\beta$--function coefficients $c_k$ and the couplant $a$.
To specify a unique result for any finite order approximant \reff{kon}
we can thus use the set \{$a,c_i$\} instead of
\{$\mu,\Lambda_{\mr{RS}},c_i$\}. At the NLO this means that instead of
the pair $\mu,\Lambda_{\mr{RS}}$ the couplant itself can serve to label
the one--parameter ambiguity! Note that neither the PMS nor the ECH
approaches fix the scale $\mu$ and the RS=\{$\Lambda_{\mr{RS}}$\}
separately, but merely their ratio, or, equivalently, just the couplant
$a$.

\section{The BLM procedure in QCD}
This method borrows its basic idea from QED, where
the renormalization of the electric charge is entirely due to the
vacuum polarization by the charged fermion--antifermion pairs.
In QCD the scale dependence of the couplant is due to other effects
as well, but the authors of \cite{BLM} suggest to fix the scale $\mu$ by
absorbing the effects of quark loops entirely in the definition of the
renormalized couplant. Let us consider the quantity \reff{r(Q)} up to
the NLO and define the class of the so--called {\em regular}
RS=$\{\Lambda_{\mr{RS}}\}$ by the condition that the expansion
coefficient $r_1$ can be written as a linear function of $n_f$
\be
r^{(2)}(Q/\mu,\mr{RS})=a(\mu,\mr{RS})\left(1+
\left[r_{10}\left(\frac{\mu}{Q},\mr{RS}\right)+
n_f r_{11}\left(\frac{\mu}{Q},\mr{RS}\right)\right]
a(\mu,\mr{RS})\right),
\label{AB}
\ee
where $r_{10},r_{11}$ are $n_f$--independent coefficients
\footnote{In this statement $\mu$ as well as the RS
(labelled by $\Lambda_{\mr{RS}}$) are held fixed and only the {\em
explicit} dependence of $r_1$ on $n_f$ is taken into account.}. The
BLM procedure fixes $\mu$ by the condition $r_{11}(\mu/Q,\mr{RS})=0$.
This implies \cite{BLM}
\begin{eqnarray}
\mu_{\mr{BLM}}(\mu,\mr{RS})
& \equiv & \mu\exp\left[3r_{11}(\mu/Q,\mr{RS})\right],
\label{muBLM} \\
r_1(\mu_{\mr{BLM}}/Q,\mr{RS})
 & = & r_{10}\left(\mu/Q,\mr{RS}\right)+
\frac{33}{2}r_{11}\left(\mu/Q,\mr{RS}\right)
\label{r1BLM}
\end{eqnarray}
and consequently
\be
r^{(2)}_{\mr{BLM}}(Q/\mu,\mr{RS})=
a(\mu_{\mr{BLM}},\mr{RS})
\left(1+\left[r_{10}\left(\frac{\mu}{Q},\mr{RS}\right)+
\frac{33}{2}r_{11}\left(\frac{\mu}{Q},\mr{RS}\right)\right]
a\left(\mu_{\mr{BLM}},\mr{RS}\right)\right).
\label{BLMres}
\ee
Note that the BLM--fixed scale $\mu_{\mr{BLM}}$ is a function of both
the ``initial'' scale $\mu$ and RS. This by itself is not surprising
as the same happens in the PMS and ECH approaches as well.
In these approaches the corresponding
couplant $a$, as well as the coefficients $r_k$, are, however,
independent of both $\mu$ and the RS, while this is not the case in
the BLM procedure. In the explicit calculation \cite{CS} Celmaster and
Stevenson have shown that for the $\Upsilon$ hadronic decay width the
coefficient $r_1$ has different values in two most frequently used RS's,
namely the \bbar{MS} and symmetric MOM RS based on the 3--gluon vertex
in the Landau gauge. Despite the fact that this simple example is
sufficient to demonstrate the basic shortcoming of the BLM procedure,
its message has largely been ignored, presumably because the dependence
of BLM results on the RS has been considered as a kind of ``residual''
dependence, less important than the dependence on the scale $\mu$. In
the rest of this section I shall demonstrate that this ``residual'' RS
dependence of the BLM procedure actually coincides with the original
scale ambiguity in a fixed RS.

To see how the BLM results depend on $\mu$ and the RS, we need to
evaluate the ratio
\be
\frac{\mu_{\mr{BLM}}(\mu,\mr{RS})}{\Lambda_{\mr{RS}}}=
\exp\left[3r_{11}\left(\frac{\mu}{Q},\mr{RS}\right)+
\ln \frac{\mu}{\Lambda_{\mr{RS}}}\right]\equiv
\exp[Z(\mu,\mr{RS})].
\label{muoverlambda}
\ee
Going from RS to RS$'$ (for fixed $\mu$), or from $\mu$ to $\mu'$
(for fixed RS), we get, exploiting \reff{r1r2},
\begin{eqnarray}
3r_{1}(\mu/Q,\mr{RS'}) & = & 3r_{1}(\mu/Q,\mr{RS})+
\left(\frac{33}{2}-n_f\right)
\ln\left(\frac{\Lambda_{\mr{RS}}}{\Lambda_{\mr{RS'}}}\right),
\label{changeRS} \\
3r_{1}(\mu'/Q,\mr{RS}) & = & 3r_{1}(\mu/Q,\mr{RS})+
\left(\frac{33}{2}-n_f\right)
\ln\left(\frac{\mu'}{\mu}\right).
\label{changemu}
\end{eqnarray}
Within the class of regular RS's we have
\begin{eqnarray}
b\ln\left(\frac{\Lambda_{\mr{RS}}}{\Lambda_{\mr{RS'}}}\right) & = &
A+Bn_f, \label{genRS} \\
Z(\mu,\mr{RS'}) & = & Z(\mu,\mr{RS})+\frac{\kappa}{b},
\label{chzRS}  \\
r_1(\mu_{\mr{BLM}}(\mu,\mr{RS'})) & = &
r_1(\mu_{\mr{BLM}}(\mu,\mr{RS}))+\kappa,\;\;\;
\kappa \equiv \frac{33}{2}B+A,   \label{chr1RS} \\
a_{\mr{BLM}}(\mr{RS'}) & = &
a_{\mr{BLM}}(\mr{RS})\left[1-\kappa a_{\mr{BLM}}(\mr{RS})\right].
\label{expansion}
\end{eqnarray}
The last two equations coincide with the transformations
\reff{change}. We see that the BLM results are unique only provided
$\kappa=33B/2+A=0$, which is equivalent to the condition that the ratio
$\Lambda_{\mr{RS}}/\Lambda_{\mr{RS'}}$ is $n_f$--independent.
This condition divides all regular RS into disjoint subclasses of RS's,
characterized by the property that for any pair of RS's from the same
subclass $\kappa=33B/2+A=0$, while for RS, RS$'$ from different
subclasses $\kappa\neq 0$. The BLM results are thus unique only within
each of these subclasses, while different subclasses lead to different
results.  To show that the latter situation arises naturally for quite
conventional RS's, let me take RS=\bbar{MS}, which is the
standard choice in all BLM papers and consider for RS$'$ the set of
MOM--based RS's, with the couplant defined via the three--gluon vertex
at the symmetric point and for general gauge fixing parameter $\alpha_G$
\cite{CG} and denote it RS($\alpha_G$). While the gauge parameter
$\alpha_G$ must also be renormalized, it can, as argued in \cite{PMS},
be considered as constant, i.e. not running. In fact
only in such case \footnote{Or for the
Landau gauge $\alpha_G=0$.} are the two lowest order $\beta$--function
coefficients $b,c$ RS--invariant constants.
Within this class of MOM--based RS's, the corresponding
$\Lambda_{\mr{MOM}}(\alpha_G)$, and thus also the couplant and expansion
coefficient $r_1$, become functions of
$\alpha_G\in(-\infty,+\infty)$. Consequently the ratio
$\Lambda_{\mr{RS}}/\Lambda_{\mr{RS'}}$ is manifestly $n_f$--dependent
and the expression for $\kappa$ in \reff{chr1RS} can be read off
eq. (18) of \cite{CG}:
\be
\kappa(\alpha_G)=
\frac{153}{48}I-\frac{9}{16}(1-I)\alpha_G-\left(-\frac{3}{8}+
\frac{I}{8}\right)\alpha^2_G-\frac{1}{16}\alpha^3_G
\label{kappa}
\ee
where $I\doteq 2.3439$. Varying the gauge fixing parameter $\alpha_G$
between $-\infty$ and $+\infty$ makes $\kappa(\alpha_G)$ to span the
same interval.  For any given $r_1$ one can thus always find such a
value of $\alpha_G$ that when $\kappa(\alpha_G)$ is used on the r.h.s.
of \reff{chr1RS}, the result is just this $r_1$. Note that at the NLO
the \bbar{MS} RS is just a special case of the MOM--based
RS($\alpha_G(\bbare{MS})$) with $\alpha_G(\bbare{MS})$ given by the
solution to the equation $\kappa(\alpha_G)=0$! So in any RS and for
any scale $\mu$, $r^{(2)}(\mu,\mr{RS})$ coincides with the result
obtained via MOM--based BLM procedure for some gauge parameter
$\alpha_G(\mu,\mr{RS})$:
\be
r^{(2)}(\mu/Q,\mr{RS})=r^{(2)}(\mu_{\mr{BLM}}/Q,\mr{RS'}
(\alpha_G(\mu,\mr{RS})))
\label{eqv}
\ee
We conclude that not only do the BLM results depend on the
chosen RS($\alpha_G$), as claimed in \cite{CS}, but
quantitatively this ambiguity is {\em completely equivalent} to
the scale ambiguity in any fixed RS.

In analogous way it can be shown that for
$n_f$--independent scale $\mu$ the BLM results
{\em are}, indeed, $\mu$--independent. Formally we could introduce
$n_f$--dependent scale $\mu$ as well and thus destroy the scale
independence of the BLM results, but as for each fixed $n_f$ the
scale $\mu$ is completely arbitrary there is no meaning in
introducing this dependence.
The situation looks differently as far as the $n_f$--dependence of the
chosen RS's is concerned, primarily because in the MOM--based RS's this
dependence arises quite naturally.

It is also simple to see why, contrary to the BLM approach, the ECH
one \cite{ECH} leads to unique results. Taking into
account \reff{r1r2} the condition $r_1=0$ implies
\be
\frac{\mu_{\mr{ECH}}(\mr{RS})}{\Lambda_{\mr{RS}}}=\exp
\frac{\rho(Q)}{b}.
\label{ratioECH}
\ee
The scale $\mu$ does not appear in \reff{ratioECH} at all and as the
r.h.s. of \reff{ratioECH} contains only RG invariants $\rho(Q)$ and
$b$, it is manifestly RS--independent. Similar considerations hold
in the PMS approach. It is thus the very essence of the BLM procedure,
i.e. the attempt to separate the coefficient $r_1$ into the
$n_f$--dependent and $n_f$--independent parts, that causes its
inherent ambiguity.

\section{The case of QED}
The results of the BLM procedure in QED can be obtained from
\reff{chzRS}--\reff{chr1RS} simply by taking into account that in QED
the corresponding $b_{\mr{QED}}=-2n_f/3$ and contains thus no term
analogous to $33/2=11N_c /2$, coming in QCD from gluon selfinteraction.
Consequently the BLM results are unique provided $\kappa=A=0$, which is
again equivalent to the condition that $\Lambda_{\mr{RS}}$ is
$n_f$--independent. In QED, however, the scale as well as
RS--dependence of the corresponding couplant
come entirely from the renormalization of fermion loops and
it is therefore natural to define the regular RS's as those
satisfying this condition. In QED and within any class of regular RS's
the BLM results are, indeed, unique.

\section{Fixing the RS by means of the ``standard'' physical process}
The idea, suggested in \cite{BLM}, is to fix the RS of the couplant
$a(\mu,\mr{RS})$ with the help of some ``standard'' physical quantity,
such as the familiar ratio
\be
 R(s)\equiv \frac{\sigma(\mr{e}^{+}\mr{e}^{-}
\rightarrow \mr{hadrons})}{\sigma(\mr{e}^{+}\mr{e}^{-}\rightarrow
\mu^{+}\mu^{-})}=3\sum_{i}e_i^2(1+r(s)),
\label{ratio}
\ee
by demanding $r(s)=a(\sqrt{s},\mr{RS})$. Note that in this way chosen
couplant is just the effective charge
corresponding to the physical quantity $r(s)$.

The main problem with this strategy is that there is no theoretical
reason to prefer one
particular physical quantity to another to serve as the ``standard''.
By appealing to some ``standard'' physical process to
fix the RS this ambiguity is transformed into the ``initial
condition'' ambiguity \cite{jaDhar,my}.
Moreover, if the ECH approach is used for the ``standard'' physical
quantity, why not to use it for the one under study?

Secondly, in writing $r(s)=a(\sqrt{s},\mr{RS})$ we have already set
the scale of the couplant $a(\mu,\mr{RS})$ to $\mu=\sqrt{s}$. Recall
that, as emphasized in Section 3, the ECH approach {\em does not}
actually fix
the scale and the RS, but determines  directly the
couplant $a_{\mr{ECH}}$, or equivalently, the ratio
$\mu/\Lambda_{\mr{RS}}$! Without specifying the scale $\mu$ in
the equation $r(s)=a_{\mr{ECH}}(\mu,\mr{RS})$ the RS is not fixed and
the reference to the ``standard'' physical quantity thus of no help.
However, to set $\mu=\sqrt{s}$ in \reff{ratio} has no theoretical
justification, in particular taking into account that the basic aim
of the BLM procedure is just to find some plausible scale fixing method!

\section{Generalization of the BLM procedure and ``genuine'' higher
order corrections}
Let me now turn to the implications of the inherent ambiguity
of the BLM procedure for some of the essential ingredients
of the recent papers \cite{BB,BBB,Neubert}, where higher
order perturbative corrections are separated into two parts:
\bit
\item  the so called ``genuine'' higher order corrections, which are
``hard to anticipate'' \cite{Neubert}, and
\item those related to the renormalization scale
dependence of the couplant, which by an ``improper''
choice of $\mu$ may become artificially large.
\eit
The starting point of these attempts is the claim \cite{BB} that
``some prescriptions may be closer to general expectations.''
Without specifying the meaning of the term ``general expectation'',
all the mentioned papers single out the BLM procedure as the
best way of fixing the scale $\mu$. One of the arguments in favour of
this conjecture is the widely held view that because the BLM approach
fixes the scale $\mu$ via the ``naive nonabelianization''
of the original QED procedure of incorporating the effects of
quark loops in the renormalized couplant, it is related to the
large order behaviour of expansion coefficients $r_k$ in \reff{r(Q)}
\cite{BB}. By generalizing the BLM procedure to higher orders the
papers \cite{BB,BBB} attempt to incorporate the effects of the
leading IR renormalon in the BLM couplant, thereby isolating the
remaining ``genuine'' higher order perturbative corrections.

The generalization
of the BLM scale--fixing procedure to arbitrary an order N suggested in
\cite{BB,BBB} is based on the generalization of the relation \reff{AB}
\footnote{As the leading order coefficient $r_0$ of \cite{BB,BBB} is
unique, it can be set to unity, as done in \reff{r(Q)}, without
losing any generality.}
\be r_n=r_{n0}+r_{n1}n_f
+\cdots +r_{nn}n_f^n.
\label{general}
\ee
Writing $r_n$ as
\be
r_n=\delta_n+(b/2)^n d_n,
\label{separace}
\ee
where $d_n$ are $n_f$--independent, absorbs the leading
$n_f$--dependence in $d_n$ and allows the authors of \cite{BB,BBB} to
define the generalized BLM scale $Q_{\mr{BLM}}^{(\mr{N})}$ at order N
via the relation \footnote{Following \cite{BB,BBB} I set in the rest
of this section $\mu=Q$.}
\be
a(Q_{\mr{BLM}}^{(N)})\equiv a(Q) M_{\mr{N}},\;\;\;M_{N}
\equiv 1+\sum_{n=1}^{N}d_n(b/2)^n a^n.
\label{Mn}
\ee
With these definitions eq. \reff{r(Q)} can be written as
\be
r(Q)=a(Q_{\mr{BLM}}^{(N)})+\sum_{n=1}^{N}\delta_{n} a(Q)^{n+1}
\label{result}
\ee
Except for the argument of the couplant in the second term of
\reff{result}, which, however, does not influence the coefficient
$\delta_1=r_1(\mu_{\mr{BLM}}/Q,\mr{RS})$, \reff{result} reduces for
$N=1$ to \reff{BLMres}. According to \cite{BB,BBB} the
coefficients $\delta_n$ represent the ``genuine'' higher
order corrections, in contrast to those included in $M_N$, which are
incorporated, via the BLM procedure, in the
leading--order couplant $a(Q_{\mr{BLM}}^{(N)})$. Moreover, in the limit
$N \rightarrow \infty$,
$Q_{\mr{BLM}}\equiv \lim_{N\rightarrow \infty}Q_{\mr{BLM}}^{(N)}$ is
claimed to be independent of the finite renormalization of the fermion
loop.

However, in QCD the renormalization of the colour charge is not
given by the gluon polarization due to fermion loops only.
Therefore the fact that each of the terms on the r.h.s. of
\reff{result} is independent of the finite renormalization of the
basic fermion loop does not imply that it is also RS--invariant and thus
unique.  In Section 3 I have shown explicitly that within the class of
MOM--based RS's both the BLM couplant $a(Q_{\mr{BLM}}^{(2)})$
and the leading coefficient $\delta_1$ of the  second term in
\reff{result} are RS--dependent. The value of the ``genuine'' NLO
correction coefficient $\delta_1$ is
a function of the the gauge--fixing parameter $\alpha_G$, and can take
any prescribed value.  The RS--dependence of $\delta_1$
automatically implies that the separation of $r(Q)$ into the two terms
in \reff{result} is ambiguous even if the BLM procedure is generalized
to an arbitrary order $N$. Recall that in perturbative expansions
of RS--independent quantities $\sum_{n=p}r_na^n$ the leading--order
coefficient $r_p$ {\em has to be unique} as there is no way how its
RS--dependence could be compensated by the RS--dependence of the
expansion parameter $a$! The only way to secure the RS--invariance of
the full sum in  \reff{result} is therefore the mutual cancellation of
this dependence between the BLM couplant $a(Q_{\mr{BLM}})$ and the
expansion describing the supposed ``genuine'' higher order
perturbative corrections.

Beyond the NLO the couplant $a$ as well as the expansion coefficients
$r_k$ become functions of the additional free parameters $c_i;i\ge 2$,
specifying the RC. This freedom implies that not only $\delta_1$, but
in fact all the coefficients $\delta_k$ are {\em completely
arbitrary}, exactly in the same way as the coefficients $r_k$ in the
general RS! In the ECH approach, for instance, they are set to zero.
The large order behaviour of the coefficients $r_k$
crucially depends on the choice of the RS and so do also all the higher
order coefficients $\delta_k$. The reason is clear: the RG
transformations mix the two terms on the r.h.s. of \reff{result} so that
we cannot attach an unambiguous meaning to each of them separately.

The fact that by an appropriate choice of the RS we can get rid of the
divergent behaviour of $r_k$ does not, of course, mean that we can in
this way solve the problem of the divergence of perturbation expansions.
The standard derivation of the asymptotic behaviour of the expansion
coefficients $r_k$ is carried out in the 't Hooft RS, in which all
higher order $\beta$--function coefficients $c_i;i\ge 2$ are set to zero
by definition. The resulting growth behaviour of the coefficients
$r_k$ at large $k$ implies via \reff{r1r2}
factorial behaviour of the invariants $\rho_k\propto k!$. This, in
turn, implies that $r_k$ are factorially divergent also in all
those RS=\{$\Lambda_{\mr{RS}},c_i$\}, where the coefficients $c_k$
define a convergent series. In the RS's where $r_k$ define convergent
series, the factorial behaviour of the invariants $\rho_k$  reappears as
the divergence of the perturbation expansion of the corresponding
$\beta$--function. Note, however, that
factorial behaviour of the coefficients $c_k$ in \reff{RG}
cannot be influenced by the choice of the scale $\mu$ and thus has
nothing to do with it. The fact that we can freely shuffle
part or the whole factorial divergence of the RG invariants $\rho_k$
between the coefficients $r_k$ and $c_k$ clearly signals that it is
impossible to relate the effects of the
factorial divergence of the invariants $\rho_k$ to the
choice of the scale, as suggested in \cite{BB,BBB, Neubert}. In other
words, there is no well--defined relation between the large order
behaviour of the coefficients $r_k$ and the choice of the
renormalization scale $\mu$. The former depends not only on $\mu$ but
also on all the parameters $c_k$, specifying the RS.

It is also fair to say that we actually do not even know how the
coefficients $c_k$ behave in the most popular RS, the \bbar{MS}.
The usual expectation is that they define a divergent series,
but there are no convincing arguments behind this conjecture.
And without this knowledge the behaviour of the expansion coefficients
$r_k$ in this RS is also an open question.

\section{Summary and conclusions}
The preceding sections demonstrate two closely related facts:
\begin{description}
\item {\bf a)}
the inherent ambiguity of the BLM scale fixing procedure, and
\item {\bf b)} the impossibility to define in a reasonably unambiguous
(i.e. RS--independent) way the ``genuine'' higher order perturbative
corrections to physical quantities.
\end{description}
The RG transformations bind inextricably the two terms in
the sum of \reff{result}. The ``genuine'' higher order corrections,
defined in \cite{BB,BBB,Neubert} by incorporating the supposed large
order behaviour of the expansion coefficients in the BLM scale, are
in fact as ambiguous as the original perturbation expansions before
the BLM scale--fixing.

The divergence of perturbative expansions can be expressed in an
unambiguous way as the statement about the factorial divergence of the
RG invariants $\rho_i$. Anything else, for instance the divergence of
the expansions in \reff{r(Q)} and/or \reff{RG}, depends on the chosen RS
and has thus no direct physical meaning. In my view the best strategy
how to proceed in
such circumstances is to follow the suggestion of \cite{PMSdiv}, i.e.
to choose at each finite order $N$ of perturbation theory a definite
renormalization prescription and investigate the limit of
finite order approximants \reff{kon}, i.e.
\be
\lim_{N\rightarrow\infty} r^{(N)}(\mu(N),\mr{RS}(N)).
\label{limit}
\ee
There is no problem to choose the
dependences $\mu(N)$ and RS($N$) in such a way that this limit is
finite, even for factorially divergent series with asymptotically
constant sign \cite{jadiv}. The question is how to do it in a
reasonably unique manner. The nonperturbative power corrections should
provide a crucial piece of information in this respect.

\vspace*{0.3cm}
\noindent
{\large \bf Acknowledgement} \\
I am grateful to P. Kol\'{a}\v{r} for many enlightening discussions and
critical comments.

\end{document}